\documentclass[twocolumn,showpacs,preprintnumbers,amsmath,amssymb,superscriptaddress]{revtex4}
\usepackage[dvipdfmx]{graphicx}
\begin{document}
\title{Photoinduced demagnetization 
and insulator-to-metal transition 
in ferromagnetic insulating BaFeO$_{3}$ thin films 
} 
\author{T. Tsuyama}
\affiliation{Institute for Solid State Physics, University of Tokyo, 
Kashiwanoha 5-1-5, Chiba 277-8581, Japan}
\affiliation{Department of Applied Physics and Quantum-Phase Electronics 
Center (QPEC), University of Tokyo, Hongo, Tokyo 113-8656, Japan}
\author{S. Chakraverty}
\affiliation{RIKEN Center for Emergent Matter Science (CEMS), Wako, Saitama 351-0198, Japan}
\affiliation{Institute of Nano Science and Technology, 
Phase - 10, Sector - 64, Mohali, Punjab, India}
\author{N. Pontius}
\affiliation{Helmholtz-Zentrum Berlin f$ \ddot{u} $r 
Materialien und Energie GmbH, Albert-Einstein-Stra\ss e 15, 12489  Berlin, Germany}
\author{C. Sch\"{u}\ss ler-Langeheine}
\affiliation{Helmholtz-Zentrum Berlin f$ \ddot{u} $r 
Materialien und Energie GmbH, Albert-Einstein-Stra\ss e 15, 12489  Berlin, Germany}
\author{H. Y. Hwang}
\affiliation{RIKEN Center for Emergent Matter Science (CEMS), Wako, Saitama 351-0198, Japan}
\affiliation{Stanford Institute for Materials and Energy Sciences, 
SLAC National Accelerator Laboratory, Menlo Park, CA 94025, USA}
\affiliation{Department of Applied Physics, 
Stanford University, Stanford, CA 94305, USA}
\author{Y. Tokura}
\affiliation{Department of Applied Physics and Quantum-Phase Electronics
Center (QPEC), University of Tokyo, Hongo, Tokyo 113-8656, Japan}
\affiliation{RIKEN Center for Emergent Matter Science (CEMS), Wako, Saitama 351-0198, Japan}
\author{H. Wadati}
\email{wadati@issp.u-tokyo.ac.jp}
\homepage{http://www.geocities.jp/qxbqd097/index2.htm}
\affiliation{Institute for Solid State Physics, University of Tokyo, 
Kashiwanoha 5-1-5, Chiba 277-8581, Japan}
\affiliation{Department of Applied Physics and Quantum-Phase Electronics 
Center (QPEC), University of Tokyo, Hongo, Tokyo 113-8656, Japan}
\date{\today}
\begin{abstract}
We studied the electronic and magnetic dynamics of 
ferromagnetic insulating BaFeO$_{3}$ thin films 
by using pump-probe time-resolved 
resonant x-ray reflectivity at the Fe $2p$ edge. 
By changing the excitation density, 
we found two distinctly different types 
of demagnetization with a clear threshold behavior. 
We assigned the demagnetization change 
from slow ($\sim$ 150 ps) to fast ($< 70$ ps) 
to a transition into a metallic state 
induced by laser excitation. 
These results provide a novel approach for locally tuning 
magnetic dynamics. 
In analogy to heat assisted magnetic recording, 
metallization can locally tune the 
susceptibility for magnetic manipulation, 
allowing to spatially encode magnetic information. 
\end{abstract}
\maketitle

Control of magnetic states by optical excitations in 
magnetically ordered materials 
has attracted considerable attention 
since the demonstration of ultrafast demagnetization 
in Ni within 1 ps, explored by time-resolved 
magneto-optical Kerr effect studies \cite{Beaurepaire1996}. 
Ultrafast demagnetization was also observed 
for other elementary ferromagnetic transition metals 
such as Co and Fe, and intermetallic alloys 
\cite{Stohr,Kirilyuk2010a}. 

Several mechanisms have been proposed 
to understand the ultrafast demagnetization.
Beaurepaire {\it et al.} proposed a phenomenological 
``three-temperature model'' 
in order to understand the ultrafast demagnetization of Ni, which
considers three interacting reservoirs of electrons, spins, and lattice, 
and suggested the importance of direct electron-spin interactions.
Since the electron, spin, and lattice systems are quite tightly 
coupled to each other in strongly correlated $3d$ transition metal
oxides, it is interesting to investigate the photoinduced 
dynamics with respect to the electronic states and magnetism 
\cite{Miyano1997,Kise2000,Cavalleri2005,Ogasawara2005,
Zhang2006,Muller2009,Koopmans2010,Okamoto2011,
Radu2011,DeJong2013,Bergeard2014,Beaud2014}. 

For this study, we chose 
fully oxidized single crystalline BaFeO$_{3}$ thin films, 
which show unusual behaviors of 
ferromagnetic and insulating properties 
with saturation magnetization and Curie temperature
of 3.2 $\mu_B/\mbox{formula unit}$ and 115 K, 
respectively \cite{Chakraverty2013}. 
The large magnetic moment of BaFeO$_{3}$ 
thin films results in quite large peak intensity of 
Fe $2p$ x-ray magnetic circular dichroism (XMCD), $\sim$ 18 \% of 
the x-ray absorption peak intensity \cite{Tsuyama2015}. 
Thus, BaFeO$ _{3} $ thin films are appropriate samples 
to carry out time-resolved magnetic circular dichroism experiments. 
Furthermore, the investigation of 
the demagnetization dynamics of insulators allows one 
to relate electronic structure to magnetic dynamics. 

\begin{figure}[ht]
\begin{center}
\includegraphics[width=7cm]{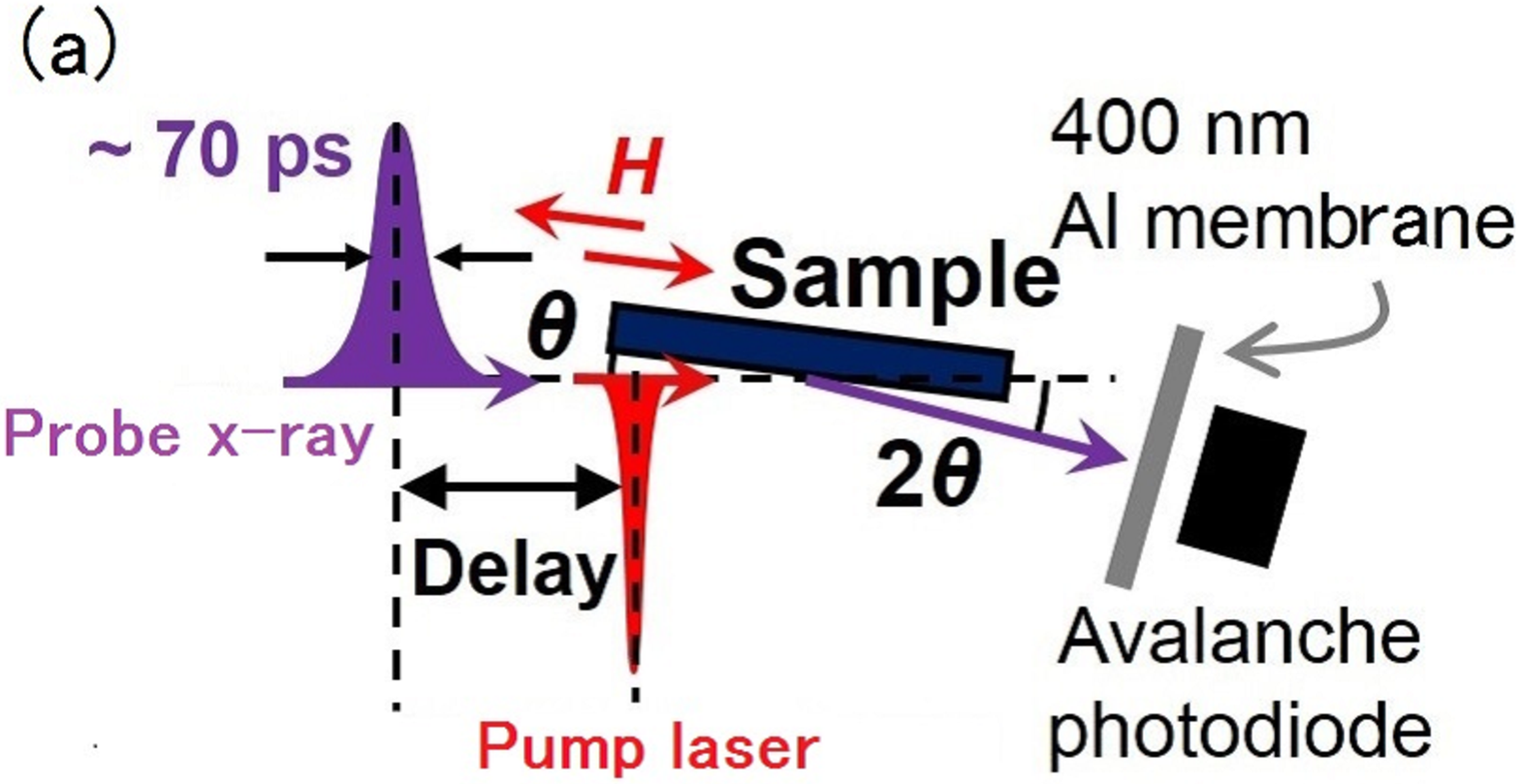}
\includegraphics[width=8cm]{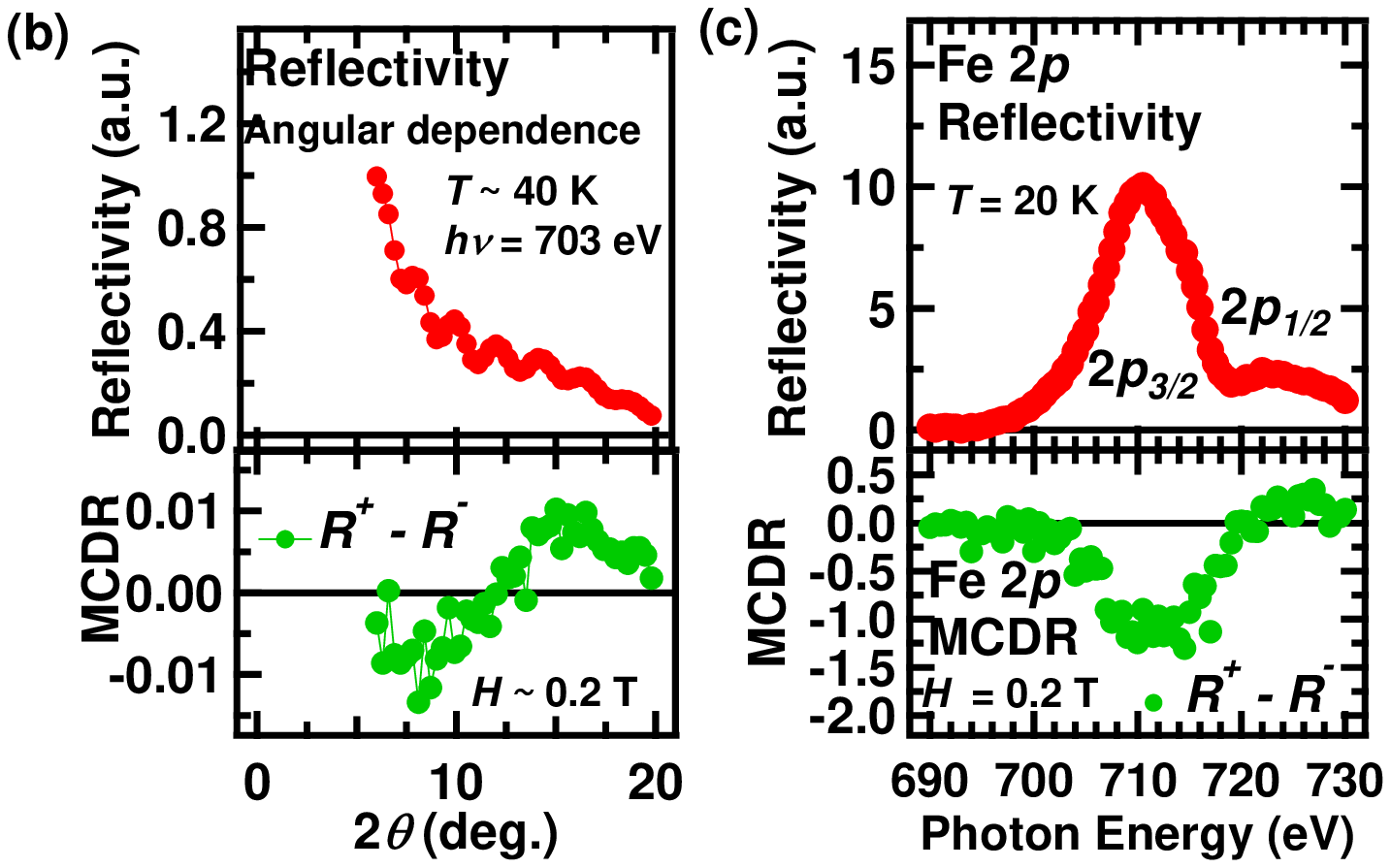}
\caption{(a) Geometry of the measurements. 
Panel (b) shows the reflectivity (top) and MCDR (bottom) 
intensity from BaFeO$_3$ thin films 
as a function of $2\theta$. Panel (c) shows the photon energy 
dependence of reflectivity (top) and MCDR (bottom).}
\label{Reflectivity}
\end{center}
\end{figure}

In order to investigate the magnetic dynamics of ferromagnetic
insulating BaFeO$_{3}$ thin films, we performed time-resolved 
reflectivity studies at the Femtospex slicing facility 
at the synchrotron radiation source  
BESSY II \cite{holl}, using circularly polarized x-ray pulses. 
Our experimental method has the advantage that, 
in one reflectivity experiment, we can probe 
electronic structure as well as magnetism. 
BaFeO$ _{2.5} $ thin films were grown on SrTiO$_{3}$ (001) 
substrate with the film thickness of 50 nm, using pulsed 
laser deposition \cite{Chakraverty2013}.
After the deposition, BaFeO$ _{2.5} $ thin films were annealed 
at 200 $ ^{\circ} $C under ozone atmosphere to obtain BaFeO$_{3}$ 
thin films \cite{Chakraverty2013}.
The quality of the thin-film samples were confirmed by x-ray
diffraction, Fe 2$p$ x-ray absorption spectroscopy, 
and Fe $2p$ core-level hard x-ray photoemission spectroscopy 
measurements by comparing cluster-model calculations, which 
found that the formal valence of Fe was $4+$ 
\cite{Chakraverty2013,Tsuyama2015}. 
The experimental geometry is shown schematically 
in Fig.~\ref{Reflectivity} (a). We used fixed circular polarization 
and created magnetic contrast by switching the direction of the 
magnetic field ($H$), which was oriented along the sample surface 
([010] direction). We recorded specular reflectivity data 
for two magnetization directions, $R^+$ and $R^-$. 
The average reflectivity $R=(R^++R^-)/2$ is a measure 
of the electronic and structural properties, while the magnetic circular 
dichroism in reflectivity (MCDR) signal 
$DR=(R^+-R^-)/2$ is a measure of the sample 
magnetization \cite{mertins}. 
A Ti-Sapphire laser ($\lambda=$ 800 nm, 
$h\nu=$ 1.55 eV) with the pulse width of $\sim 50$ fs 
was employed as a pump laser with $\pi$ polarization. 
The spot size of the pump laser was 
$\sim$ 0.40 mm (horizontal) $\times$ 0.25 mm (vertical), 
and that of the probe x-ray was 
$\sim$ 0.1 mm $\times$ 0.1 mm. 
The repetition rate of the time-resolved measurement 
was 3 kHz, limited by the frequencies of the pump laser. 
The pumped and unpumped signals were obtained alternatively. 
The time resolution was 70 ps, corresponding 
to the pulse length of the probe x-ray. 

The angular dependences of reflectivity and MCDR are 
shown in Fig.~\ref{Reflectivity} (b) 
for a photon energy of 703 eV. 
The angular dependence of reflectivity show oscillating structures, 
attributed to the interference between x-rays reflected from 
the surface and interface of the thin-film sample.
It enables us to estimate the film thickness of the sample to be 
(48 $\pm$ 1) nm, in good agreement with the evaluation of 
$\sim$ (50 $\pm$ 1) nm by reflection high-energy electron diffraction.
The oscillating structures in the angular dependence of reflectivity 
also imply a probing depth larger than the film thickness.
MCDR also shows an angular dependence with sign reversal 
as shown in Fig.~\ref{Reflectivity} (b).
We fixed 2$\theta$ to be $\sim$ 15$^{\circ}$, which maximizes the MCDR 
to about 12 \% of the average reflectivity. 
Figure \ref{Reflectivity} (c) shows 
the photon-energy dependence of reflectivity and MCDR. 
Although the photon-energy dependence of MCDR 
is different from the circular dichroism obtained 
from x-ray absorption spectra \cite{Tsuyama2015}, 
it is a suited measure of the sample magnetization \cite{mertins}. 
We fixed the photon energy to be $h\nu$ = 714 eV, and the time evolution 
of the intensities of reflectivity as well as XMCD at this photon energy are 
traced in order to investigate the electronic and magnetic dynamics 
of BaFeO$_{3}$ thin films.

\begin{figure}[ht]
\begin{center}
\includegraphics[width=8cm]{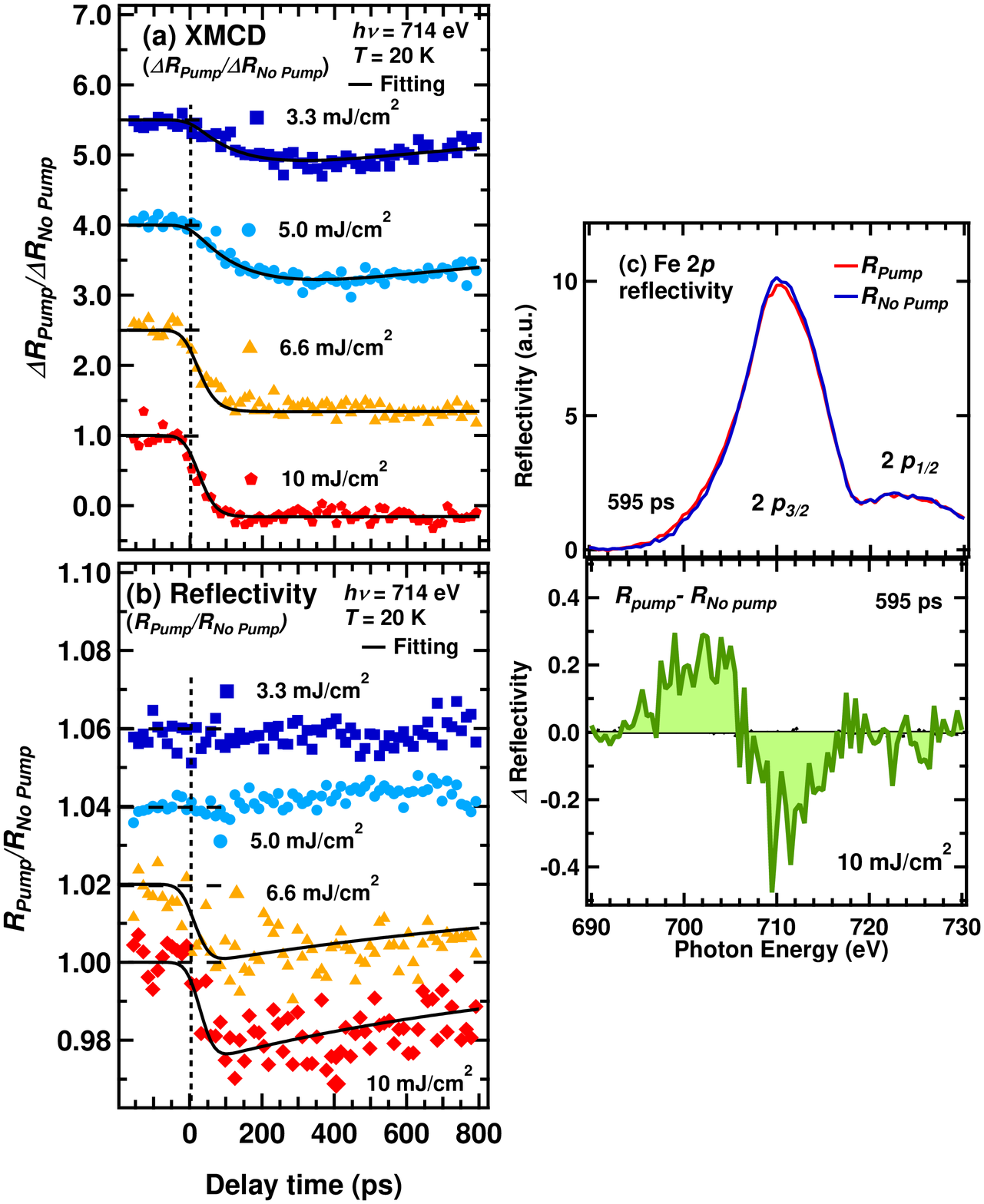}
\caption{Time evolution of (a) XMCD intensity and (b) reflectivity 
of BaFeO$_{3}$ thin films for various pump laser fluence. 
All the curves, except for the case of 10 mJ/cm$^2$, 
are shifted upward for clarity. 
(c) The reflectivity spectra with and without the pump effect (top). 
The difference of the spectra (bottom).}
\label{TrXMCD}
\end{center}
\end{figure}
	  
Figure~\ref{TrXMCD} (a) shows the time evolution of the MCDR intensities 
for different pump fluence. 
The vertical axis shows the excited MCDR intensities normalized 
by the unpumped signal 
($i.e.$ $\it{\Delta}R_{Pump}$/$\it{\Delta}R_{No Pump}$). 
Here, the subscript of Pump and No Pump denote the signals with and 
without the laser excitations, respectively.
The MCDR intensities decrease after the incidence 
of the pump laser at $t=0$. 
The time evolution of the MCDR intensity show different behaviors 
with the change of the pump fluence ($F$). 
When $F$ is smaller than 5.0 mJ/cm$^{2}$, the demagnetization
time is relatively slow and 
magnetization recovery sets in after about 400 ps. 
When the pump $F$ is larger than 6.6 mJ/cm$^{2}$, on the other hand, 
the demagnetization time is quite fast and no recovery of the
magnetization can be observed  within the first 800 ps. 
We assign the different behavior of the demagnetization dynamics 
to a laser-induced insulator-to-metal transitions 
for $F\geq$ 6.6 mJ/cm$^{2}$, as discussed in the following. 

We show the time evolution of the intensity of the average reflectivity 
in Fig.~\ref{TrXMCD} (b), which allows us 
to investigate the electronic dynamics. 
The vertical axis shows the excited reflectivity intensities normalized 
by those without excitation ($i.e.$ $R_{Pump}$/$R_{No Pump}$). 
No pump effects were observed for $F\leq$  5.0 mJ/cm$^{2}$ 
with our time resolution of 70 ps.
Pump effects, on the other hand, were clearly observed 
for $F\geq$ 6.6 mJ/cm$^{2}$. 
They occur via a transfer of spectral weight from the maximum of 
the Fe $2p_{3/2}$ resonance (710 eV) towards lower excitation energies 
as evidenced by comparing the spectra with and without laser 
in Fig.~\ref{TrXMCD} (c). The transfer to lower energies 
is particularly evident in the difference plot in the lower frame. 
A similar behavior with spectral weight transfers above a fluence 
threshold was observed in time-resolved x-ray absorption 
spectroscopy from VO$_2$ \cite{Cavalleri2005} 
and interpreted as a laser induced insulator-to-metal transition. 
Since BaFeO$_3$ thin films are near the phase boundary 
of a metal-insulator transition, we assign the spectral weight transfer 
in this material to a similar mechanism. 

In order to discuss the results quantitatively,  
we fitted them using exponential functions for decay and 
recovery (solid lines in Fig.~\ref{TrXMCD} (a) and (b)); 
$\tau_{decay}$ and $\tau_{recovery}$ denote 
the corresponding time constants. 
To take temporal resolution into account, 
the fitting function was convoluted by a Gaussian 
with a full width at half maximum of $\tau_{reso}$ = 70 ps. 
The parameters extracted from the experimental delay scans 
are summarized in TABLE~\ref{table1}. 

\begin{table}[ht]
\begin{flushleft}
\caption{Fitting parameters of 
the time evolution of 
MCDR and reflectivity intensity in Fig.~\ref{TrXMCD}. 
We fixed $\tau_{reso}$ = 70 ps.}
\begin{center}
\begin{tabular}{c|cc|cc}
\hline\hline
$F$ & $\tau_{decay,MCDR}$ & $\tau_{recovery,MCDR}$ & 
$\tau_{decay,ref}$ & $\tau_{recovery,ref}$ \\ 
$[\mbox{mJ/cm}^2]$ & [ps] & [ps] & [ps] & [ps] \\
\hline
$3.3$ & $140$ & $1000$ & $-$ & $-$ \\
$5.0$ & $150$ & $1200$ & $-$ & $-$ \\
$6.6$ & $29$ & $-$ & $16$ & $1200$ \\
$10$ & $23$ & $-$ & $9.0$ & $1000$ \\
\hline\hline
\end{tabular}
\label{table1}
\end{center}
\end{flushleft}
\end{table}

\begin{figure}[ht]
\begin{center}
\includegraphics[width=7cm]{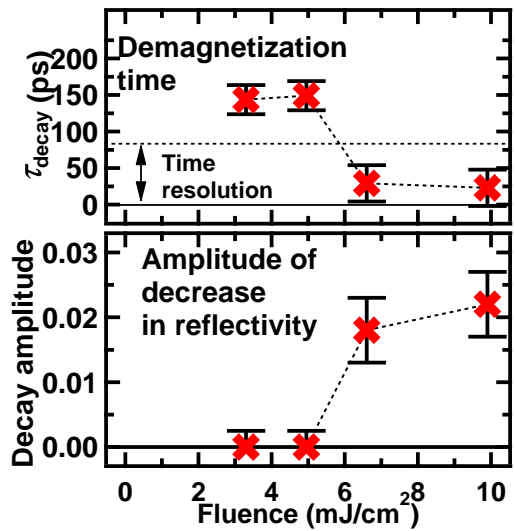}
\caption{The demagnetization time $\tau_{decay}$ (top), 
determined from Fig.~\ref{TrXMCD}, and the magnitude 
of decay of reflectivity due to the pump (bottom) are shown.}
\label{Summary}
\end{center}
\end{figure}

The quantitative analysis confirms the existence of two different kinds 
of dynamics for high and low fluences respectively. 
We show the demagnetization time, $\tau_{decay}$ determined 
by the fitting functions in the top panel of Fig.~\ref{Summary}.
The demagnetization time is estimated to 
be $\tau_{decay}$ $\sim$ 150 ps for the weaker excitations 
($F\leq $ 5.0 mJ/cm$^{2}$), and demagnetization faster 
than our temporal resolution was 
observed for the strong pump fluence ($F\geq $ 6.6 mJ/cm$^{2}$).
In the bottom panel of Fig.~\ref{Summary}, we show the 
amplitude of the reflectivity change, which is a measure of 
the spectral weight transfer indicative 
of the insulator-to-metal transition, 
as a function of laser fluence. 
From the clear threshold behavior of this quantity 
we deduce that a sufficiently high density of 
excited carriers reduces 
electron-electron and electron-phonon interactions, 
which are the origins of the insulating properties 
in BaFeO$_{3}$ thin films \cite{Chakraverty2013, Tsuyama2015}. 

We then conclude that the drastic changes of in the demagnetization 
dynamics with laser fluence are due to the photoinduced 
insulator-to-metal transition as shown by the time-resolved 
reflectivity measurement, since the electronic structure 
significantly influences the demagnetization 
dynamics \cite{Kise2000,Zhang2006,Muller2009,Kirilyuk2010a}. 
Half metals and ferromagnetic insulators usually show relatively 
slow demagnetization in the range of from 100 ps to 1000 ps, 
while itinerant ferromagnets (with low spin polarization) 
can show ultrafast demagnetization 
\cite{Kise2000,Zhang2006,Muller2009,Kirilyuk2010a}. 
Within common models for ultrafast magnetic dynamics, these differences 
in demagnetization dynamics are explained 
by the differences of strength of coupling between 
electronic and spin reservoirs. 
Ferromagnetic metals with low spin polarization, 
such as Ni and Fe, have strong coupling between 
electron and spin reservoirs because the electronic structure 
of ferromagnetic metals can accommodate spin-flip excitations with 
quasiparticle scatterings, resulting in the ultrafast 
demagnetization due to the rapid increase of spin temperature 
via the electron reservoir.
Half metals and ferromagnetic insulators, on the other hand, 
do not have strong coupling between the reservoirs.
This is because the spin-flip excitations are significantly 
suppressed after the rapid increase of electron temperature 
in their electronic structure, resulting in slow increase 
of spin temperature via heating of the lattice.

Since undisturbed BaFeO$_{3}$ thin films are ferromagnetic insulators, 
the spin scattering after the electronic excitations should be suppressed.
The slow demagnetization time  of $\sim$ 150 ps in BaFeO$_{3}$ thin
films with small pump fluence ($F\leq$ 5.0 mJ/cm$^{2}$) can be explained 
by heating via the lattice. 
The large pump fluence ($F\geq$ 6.6 mJ/cm$^{2}$), however, induces 
an insulator-to-metal transition in BaFeO$_{3}$ thin films quite rapidly, 
which results in the unusually fast demagnetization in BaFeO$_{3}$ 
thin films for a ferromagnetic insulator. 

\begin{figure}[ht]
\begin{center}
\includegraphics[width=7cm]{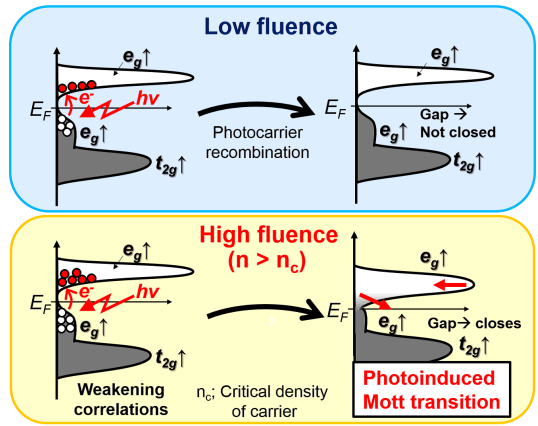}
\caption{Mechanism of insulator-to-metal transition 
induced by the strong laser excitation.}
\label{IM}
\end{center}
\end{figure}

When the pump fluence is weaker than 5.0 mJ/cm$^{2}$, 
magnetizations in BaFeO$ _{3} $ thin films recover with the time
constant of $\tau_{recovery}$ $\sim$ 1000 ps.
The time scale of  $\sim$ 1000 ps can be assigned to heat diffusion 
needed to cool the sample below the magnetic ordering temperature 
after electron, lattice, and spin systems have reached 
thermal equilibrium. 
Remarkably, the time-resolved reflectivity change 
for strong pump fluence also shows a recovery 
on this time scale of $\tau_{recovery}$ $\sim$ 1000 ps, 
indicating that also here heat diffusion is the relevant mechanism. 
This latter observation is quite notable, 
because equilibrium between the electron and lattice temperature 
should be reached within 1 ps. The slow reopening of the band-gap 
on time scales of $\sim$ 1000 ps shows that for high excitation
densities we drive the system into a metastable state. 
As a mechanism for the long life time of the metallic state, 
we consider that hot carriers, generated by the quasiparticle 
scattering and closing the band gap, prevent it from opening again 
by reducing electron-electron and 
electron-phonon interactions \cite{Okamoto2011}. 
This process is schematically shown in Fig.~\ref{IM}. 

In conclusion, we investigated the electronic and magnetic 
dynamics by time-resolved reflectivity and MCDR measurement 
on BaFeO$_{3}$ thin films.
When the pump laser fluence is smaller than 5.0 mJ/cm$^{2}$, 
relatively slow demagnetization of $\tau_{decay}$ $\sim$ 150 ps was 
observed, due to the insulating properties of the ground state 
in BaFeO$_{3}$ thin films without any changes in Fe $2p$ x-ray 
reflectivity.
When the pump laser fluence is stronger than 6.6 mJ/cm$^{2}$, on the 
other hand, rapid changes in Fe $2p$ x-ray reflectivity are 
observed, which is attributed to a transition into a metallic state, 
resulting in an unusually fast demagnetization 
with $\tau_{decay} < 70$ ps. 
Since BaFeO$_{3}$ thin films are near the phase boundary 
of a metal-insulator transition, the insulating phase is 
quite sensitive to carrier density. Thus, the origin of the 
insulator-to-metal transition is a photoinduced Mott transition 
into a metastable state stabilized by screened electron-electron 
and electron-phonon interactions. 
Our findings indicate a mechanism for tuning magnetic dynamics 
in correlated materials, which resembles heat-assisted 
magnetic switching in metallic magnets. By creating 
a sufficiently high excitation density, spin flip scattering channels 
open up which increase the spin systems susceptibility 
to external manipulation. 

This research was supported by the Japan Society
the Promotion of Science (JSPS) through the Funding Program
for World-Learning Innovative R\&D on Science and Technology
(FIRST program), JSPS Giant-in-Aid for Scientific
Research, and by Grant for Basic Science Research Projects
from the Sumitomo Foundation. This work was also partially
supported by the Ministry of Education, Culture, Sports,
Science and Technology of Japan (X-ray Free Electron Laser
Priority Strategy Program).
H. Y. H. acknowledges support by the Department of Energy, 
Basic Energy Sciences, Materials Science and Engineering 
Division, under Contract No. DE-AC02- 76SF00515.


\begin{thebibliography}{24}
\bibitem{Beaurepaire1996} E. Beaurepaire, J.-C. Merle, A. Daunois, and J.-Y. Bigot, Phys. Rev. Lett. $ \bf{76} $, 4250 (1996).
\bibitem{Kirilyuk2010a} A. Kirilyuk, A. V. Kimel, and T. Rasing, Rev. Mod. Phys. $ \bf{82} $, 2731 (2010).
\bibitem{Stohr} J. St\"{o}hr and H. C. Siegmann, 
{\it Magnetism} (Springer, Berlin, 2006). 
\bibitem{Miyano1997} K. Miyano, T. Tanaka, Y. Tomioka, and Y. Tokura, Phys. Rev. Lett. $ \bf{78} $, 4257 (1997).


\bibitem{Kise2000} T. Kise, T. Ogasawara, M. Ashida, Y. Tomioka, 
Y. Tokura, and M. Kuwata-Gonokami, Phys. Rev. Lett. 
$\bf{85}$, 1986 (2000).

\bibitem{Cavalleri2005} A. Cavalleri, M. Rini, H. H. W. Chong, 
S. Fourmaux, T. E. Glover, P. A. Heimann, J. C. Kieffer, and 
R. W. Schoenlein, Phys. Rev. Lett. $\bf{95} $, 067405 (2005).

\bibitem{Ogasawara2005} 
T. Ogasawara, K. Ohgushi, Y. Tomioka, K. S. Takahashi, 
H. Okamoto, M. Kawasaki, and Y. Tokura, 
Phys. Rev. Lett. $ \bf{94} $, 087202 (2005).

\bibitem{Zhang2006} 
Q. Zhang, A. V. Nurmikko, G. X. Miao, 
G. Xiao, and A. Gupta, 
Phys. Rev. B  $ \bf{74} $, 064414 (2006).

\bibitem{Muller2009} G. M. Muller, J. Walowski, M. Djordjevic, 
G.-X. Miao, A. Gupta, A. V Ramos, K. Gehrke, V. Moshnyaga, 
K. Samwer, J. Schmalhorst, A. Thomas, A. Hatten, G. Reiss, 
J. S. Moodera, and M. Monzenberg,  Nat. Mater. $\bf{8}$, 56 (2009).

\bibitem{Koopmans2010} B. Koopmans, G. Malinowski, F. Dalla Longa, 
D. Steiauf, M. Fahnle, T. Roth, M. Cinchetti, and M. Aeschlimann, 
Nat. Mater. $\bf{9}$, 259  (2010). 


\bibitem{Okamoto2011} H. Okamoto, T. Miyagoe, K. Kobayashi, H. Uemura,
H. Nishioka, H. Matsuzaki, A. Sawa, and Y. Tokura, 
Phys. Rev. B  $ \bf{83} $, 125102 (2011).

\bibitem{Radu2011} I. Radu, K. Vahaplar, C. Stamm, T. Kachel, 
N. Pontius, H. A. Durr, T. A. Ostler, J. Barker, R. F. L. Evans, 
R. W. Chantrell, A. Tsukamoto, A. Itoh, A. Kirilyuk, T. Rasing, and A. V. Kimel,  
Nature $\bf{472}$, 205 (2011). 

\bibitem{DeJong2013} S. de Jong, R. Kukreja, C. Trabant, N. Pontius, 
C. F. Chang, T. Kachel, M. Beye, F. Sorgenfrei, C. H. Back, 
B. Br$\ddot{a}$uer, W. F. Schlotter, J. J. Turner, O. Krupin, 
M. Doehler, D. Zhu, M. A. Hossain, A. O. Scherz, D. Fausti, 
F. Novelli, M. Esposito, W. S. Lee, Y. D. Chuang, D. H. Lu, R. G. Moore, 
M. Yi, M. Trigo, P. Kirchmann, L. Pathey, M. S. Golden, M. Buchholz, 
P. Metcalf, F. Parmigiani, W. Wurth, A. Fohlisch, 
C. Sch\"{u}\ss ler-Langeheine, and H. A. D$ \ddot{u} $rr,  
Nat. Mater. $\bf{12}$, 882 (2013).

\bibitem{Bergeard2014} N. Bergeard, V. Lopez-Flores, 
V. Halt, M. Hehn, C. Stamm, N. Pontius, E. Beaurepaire, 
and C. Boeglin, Nat. Commun. $\bf{5}$, 3466 (2014). 

\bibitem{Beaud2014} P. Beaud, A. Caviezel, S. O. Mariager, 
L. Rettig, G. Ingold, C. Dornes, S.-W. Huang, J. A. Johnson, 
M. Radovic, T. Huber, T. Kubacka, A. Ferrer, H. T. Lemke, 
M. Chollet, D. Zhu, J. M. Glownia, M. Sikorski, A. Robert, 
H. Wadati, M. Nakamura, M. Kawasaki, Y. Tokura, S. L. Johnson, 
and U. Staub, Nat. Mater. $\bf{13}$, 923 (2014).

\bibitem{Chakraverty2013} S. Chakraverty, T. Matsuda, N. Ogawa, 
H. Wadati, E. Ikenaga, M. Kawasaki, Y. Tokura, and H. Y. Hwang, 
Appl. Phys. Lett. $\bf{103}$, 142416 (2013).

\bibitem{Tsuyama2015} T. Tsuyama, T. Matsuda, S. Chakraverty, 
J. Okamoto, E. Ikenaga, A. Tanaka, T. Mizokawa, H. Y. Hwang, 
Y. Tokura, and H. Wadati, Phys. Rev. B $\bf{91}$, 115101 (2015). 

\bibitem{holl} K. Holldack, J. Bahrdt, A. Balzer, 
U. Bovensiepen, M. Brzhezinskaya, A. Erko, A. Eschenlohr, 
R. Follath, A. Firsov, W. Frentrup, L. Le Guyader, T. Kachel, 
P. Kuske, R. Mitzner, R. Muller, N. Pontius, T. Quast, I. Radu, 
J.-S. Schmidt, C. Sch\"{u}\ss ler-Langeheine, M. Sperling, 
C. Stamm, C. Trabant and A. Fohlisch, 
J. Synchrotron Rad. {\bf 21}, 1090 (2014). 

\bibitem{mertins}
H.-Ch. Mertins, D. Abramsohn, A. Gaupp, F. Schafers, 
W. Gudat, O. Zaharko, H. Grimmer, and P. M. Oppeneer, 
Phys. Rev. B {\bf 66}, 184404 (2002). 


\end{thebibliography}
\end{document}